\documentclass[twocolumn,showpacs,prl,superscriptaddress]{revtex4}

\usepackage{graphicx}
\usepackage{amssymb}

\begin{document}
\newcommand{\qco}{$q_{\rm co}$}
\newcommand{\OMco}{$\Omega_{\rm co}$}
\newcommand{\OMbp}{$\Omega_{\rm bp}$}
\newcommand{\dibo}{Li$_2$O-2B$_2$O$_3$}

\title{Acoustic damping in \dibo{} glass observed by inelastic x-ray and optical Brillouin scattering}

\author{B. Ruffl\'e}
\author{G. Guimbreti\`ere}
\author{E. Courtens}
\author{R. Vacher}
\affiliation{Laboratoire des Collo\"{\i}des, Verres et Nanomat\'eriaux, UMR 5587 CNRS, Universit\'e Montpellier II, F-34095 Montpellier Cedex 5, France}
\author{G. Monaco}
\affiliation{European Synchrotron Radiation Facility, Bo\^{\i}te Postale 220, F-38043 Grenoble, France}

\date{\today}

\begin{abstract}
The dynamic structure factor of lithium-diborate glass has been measured at several values of the momentum transfer $Q$ using high resolution inelastic x-ray scattering. Much attention has been devoted to the low $Q$-range, below the observed Ioffe-Regel crossover \qco{}$\simeq$ 2.1 nm$^{-1}$. We find that below \qco{}, the linewidth of longitudinal acoustic waves increases with a high power of either $Q$, or of the frequency $\Omega$, up to the crossover frequency \OMco{} $\simeq$ 9 meV that nearly coincides with the center of the boson peak. This new finding strongly supports the view that resonance and hybridization of acoustic waves with a distribution of rather local low frequency modes forming the boson peak is responsible for the end of acoustic branches in strong glasses. Further, we present high resolution Brillouin light-scattering data obtained at much lower frequencies on the same sample. These clearly rule out a simple $\Omega^2$-dependence of the acoustic damping over the entire frequency range.
\end{abstract}

\maketitle

\section{Introduction}

The nature of acoustic excitations at high frequency in glasses is still a central issue in the physics of disordered materials. The understanding of these collective vibrations is far from the level reached for crystals. At long wavelengths $\lambda = 2\pi/q$, or low frequencies $\Omega$, the acoustic excitations are well approximated in glasses by propagating plane-waves with a linear dispersion $\Omega(q)$. The attenuation is mainly due to tunneling or thermally activated relaxation of defects \cite{Hunklinger1976} and anharmonicity \cite{Vacher1981}. In a momentum conserving scattering experiment, $q$ is given by the geometrically determined momentum transfer $Q$, $q=Q$. Increasing the scattering vector $Q$, the simple relation $\Omega(Q) \propto Q$ starts to deviate from linearity in the THz region. In this high-frequency region, several glass-specific damping mechanisms are expected, such as Rayleigh scattering \cite{Graebner1986} or resonance with the opticlike modes that give rise to the boson peak \cite{Buchenau1992,Gurevich2003}. These processes lead to a very rapid increase of the phonon linewidth $\Gamma\propto\Omega^\alpha$, with a high power $\alpha\simeq4$, up to a Ioffe-Regel crossover frequency \OMco{} \cite{Ioffe1960} where the energy mean free path of the acoustic excitations equals half their wavelength, $\ell = \lambda/2$. The latter relation leads to the crossover criterion $\Gamma \simeq \Omega/\pi$. In all glasses where the boson peak is sufficiently strong, we have shown that \OMco{} nearly coincides with this peak, \OMco{} $\simeq$ \OMbp{} \cite{Rat1999, Ruffle2003, Courtens2003, Ruffle2005}. Using the soft-potential model, a similar relation was already obtained in \cite{Parshin2001}. Above the crossover, the acoustic modes do not anymore possess a well defined wave vector $q$, strictly implying the end of the acoustic ``branch'' $\Omega(q)$. The acoustic excitations merge then with the boson peak \cite{Foret2002, Ruffle2005}. These findings provide a natural explanation for the low-temperature thermal anomalies of glasses, in particular for the plateau found in the thermal conductivity $\kappa(T)$ around 10 K \cite{Zeller1971}. 

Despite these experimental evidences, a general consensus on the exact nature of sound excitations at THz frequencies is still missing as can be inferred from highly debated sessions at the 5$^{\rm th}$ IDMRCS conference \cite{ThisProc2005} on that topic. In this paper, we present part of the recent IXS data we obtained in \dibo{} at $T$ = 573 K. We mainly focus on the low-$Q$ range where a strong increase of the Brillouin linewidth is observed, {\em i.e.} below the crossover \qco{}. We demonstrate that (i) the measured inelastic intensity is {\em entirely} due to the longitudinal acoustic excitation; (ii) $\Gamma \propto \Omega^\alpha$, or $\Gamma \propto Q^\alpha$, with $\alpha \simeq 4$ up to a Ioffe-Regel crossover \qco{} $\simeq$ 2.1 nm$^{-1}$; (iii) above the crossover, the measured spectral shapes cannot anymore be described by a damped harmonic oscillator. Further, high resolution Brillouin light-scattering data obtained at lower frequencies reveal an acoustic damping $\Gamma \propto \Omega$ in \dibo{}. This shows that the strength of the other important broadening mechanisms, thermally activated relaxations and anharmonicity, {\em strongly depends both on temperature and on the particular glass}. There is no physical reason to search for a single universal law $\Gamma \propto Q^2$, at least not below the crossover.

\section{Experimental details}

The inelastic x-ray experiment was performed at the high-resolution inelastic beam-line ID16 at the European Synchrotron Radiation Facility in Grenoble. The monochromatic beam is obtained by a double-crystal pre-monochromator and a high-resolution monochromator operating at the Si(11,11,11) reflection in the near-backscattering geometry. The scattered beam is collected by a spherical silicon crystal analyzer, also operating at the Si(11,11,11) back reflection. A five analyzer set-up allows the simultaneous recording of spectra at five $Q$-values spaced by 3 nm$^{-1}$, each with a $Q$-resolution of $\pm$ 0.18 nm$^{-1}$. Energy scans are performed by varying the relative temperature of the monochromator and analyzer crystals. The incident x-rays have an energy of 21.748 eV and their intensity is 2 $\times$ 10$^8$ photons/s. The total energy resolution, measured from the elastic scattering of a Plexiglas sample at the maximum of its structure factor ($Q$ = 10 nm$^{-1}$), was 1.5 meV full width at half maximum (FWHM). Each scan has taken about 210', and each final spectrum has been obtained from the average of 2 to 3 scans depending on the sample temperature. The data have been normalized to the incident beam intensity. The small constant electronic background has been measured for each detector. 

The sample was kindly provided by Dr. Matic from Chalmers University of Technology, G\"oteborg, Sweden. The sample length of $\sim$ 8 mm was chosen to be comparable to the x-ray photoabsorption length, resulting in negligible multiple scattering effects. The sample was placed into a cylindrical home-made oven, sealed at both ends by thin aluminized Mylar windows. The background originating from the sample environment was found to be completely negligible. Measurements have been taken at room temperature, and at a higher temperature, 573 K, to increase the Brillouin signal by the Bose factor. IXS spectra were recorded at $Q$ ranging from 1 to 14.7 nm$^{-1}$. Furthermore, a high-$Q$ set of spectra, between 20 and 29 nm$^{-1}$, has been obtained to characterize the boson peak.  

The optical Brillouin scattering spectra have been measured with a high resolution spectrometer that consists in a four-pass planar Fabry-Perot (PFP) monochromator followed by a spherical Fabry-Perot (SFP) analyzer. This is a modern version of an instrument already described in \cite{Vacher1980}. A single mode Ar$^+$ laser operating at 514.5 nm is used to excite the Brillouin spectrum. The free spectral range of the SFP was 1.497 GHz and its finesse 60, leading to a resolution of 25 MHz FWHM. The sample was the one used for the IXS  experiment. The Brillouin frequency shift and the Brillouin linewidth have been obtained as a function of temperature and scattering angle. The analyzed $Q$-values range from 0.09 to 0.038 nm$^{-1}$.

\section{Results}

\begin{figure}
 \includegraphics[width=8.5cm]{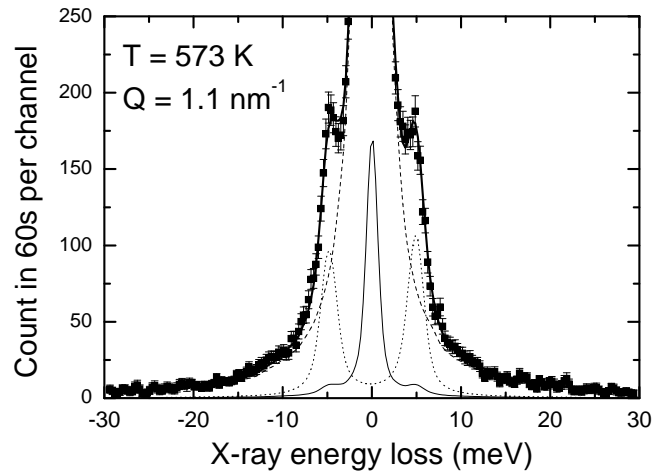}
\caption{The IXS spectrum of \dibo{} at $Q$ = 1.1 nm$^{-1}$ and T = 573 K (symbols) together with the result of fitting with the sum of a delta function and a DHO, convoluted with the resolution function (solid curve). The dashed curve shows the resolution function whereas the convoluted inelastic part is represented by the dotted curve. Also shown is the full spectrum $\times$ 1/20 (thin central line).}
\label{IXSwholespectrum}
\end{figure}

Fig. \ref{IXSwholespectrum} shows the high temperature spectrum obtained at the smallest usable $Q$-value. An elastic peak plus a Brillouin doublet, corresponding to the longitudinal acoustic (LA) excitation, are clearly observed. The fitting procedure determines the energy position, $\Omega(Q)$, and the width, $\Gamma(Q)$, of the inelastic excitation. All the measured spectra have been adjusted to a delta function for the elastic scattering plus a damped harmonic oscillator (DHO), convoluted with the instrumental response, taking into account the frequency spread produced by the collection aperture.

\begin{figure}
 \includegraphics[width=8.5cm]{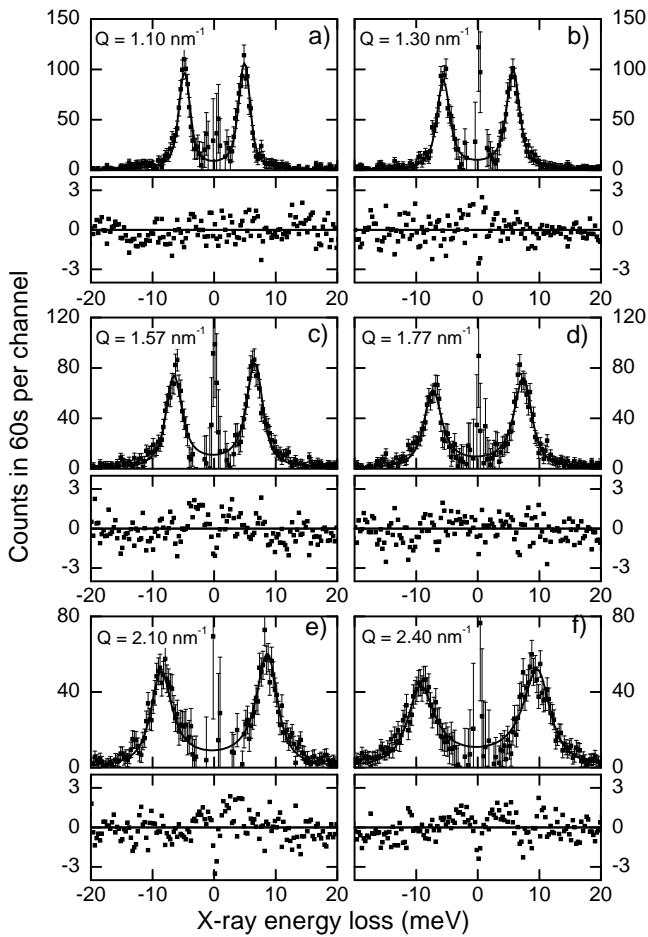}
\caption{$Q$-dependence of the IXS inelastic spectra of \dibo{} at 573 K and their fits explained in the text. For this presentation, the {\em adjusted} elastic part is subtracted from both the experimental data and the global fit of each spectrum. The lines show the DHO adjustment resulting from the global fit. The bottom panel of each subplot reports the difference between the experimental data and the fit normalized to the standard deviation at each point.}
\label{IXSspectra}
\end{figure}

Fig. \ref{IXSspectra} shows the $Q$-dependence of the inelastic part of the IXS spectra at 573 K in the low momentum transfer region. The small constant electronic background is already subtracted. For clarity, the {\em adjusted elastic peak} has been subtracted from both the data and the solid line, {\em maintaining the original error bars}. This presentation just makes the Brillouin doublet and the quality of the fit well apparent without affecting the fitting procedure in any way. For $Q$ = 1.1 nm$^{-1}$, the Brillouin linewidth observed in Fig. \ref{IXSspectra}(a) is mostly instrumental. A rapid increase of the real broadening of the longitudinal acoustic mode is clearly seen at higher $Q$-values. In Fig. \ref{IXSspectra}(e), for which $Q \sim$ \qco{}, the real broadening becomes the dominant part. The bottom panel of each subplot in Fig. \ref{IXSspectra} reports the difference between the experimental data and the fit normalized to the standard deviation at each point. This clearly emphasizes the high quality of the fit up to \qco{}. Above \qco{}, one can already detect in Fig. \ref{IXSspectra}(f) that the DHO lineshape starts to deviate from the measured signal as discussed below.

\begin{figure}
 \includegraphics[width=8.5cm]{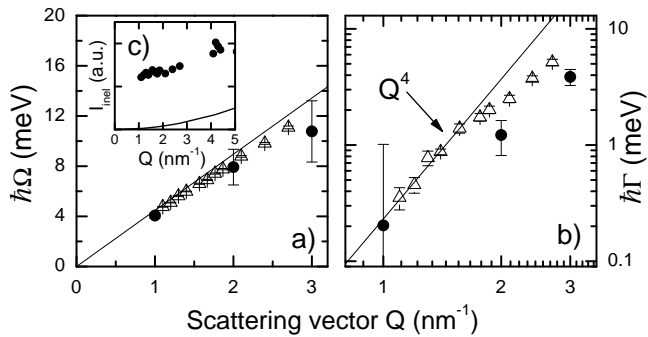}
\caption{The parameters of the DHO fit at 573 K. The three full dots at 700 K are from \cite{Matic2001}. The slope of the line in (a) is the BLS velocity measured independently at 573 K. The line of slope 4 in the log-log plot (b) is drawn through the first five points. The inset (c) shows the integrated inelastic intensities at 573 K, corrected for the oxygen-ion form factor, with a line $\propto Q^2$ scaled to the $Q$ = 29.4 nm$^{-1}$ value.}
\label{DhoParameters}
\end{figure}

The $Q$-dependence of the parameters extracted from the DHO fits of the IXS spectra are displayed in Fig. \ref{DhoParameters} for $T$ = 573 K. The parameters at 300 K are quite similar \cite{Ruffle2005}. At the lowest $Q$-value, the energy position indicates that the longitudinal acoustic modes propagate with the same sound velocity found at lower frequencies in BLS experiments at this temperature. The data reveal an almost linear dispersion which slowly  departs from linearity as $Q$ increases. Fig. \ref{DhoParameters}(b) shows however that the full width $\Gamma$ increases very rapidly for the smallest $Q$ values. Indeed, a power law $\sim Q^4$ is found for the five lowest points. The crossover \qco{}$\simeq 2.1$ nm$^{-1}$, with \OMco$\simeq 9$ meV, is determined from the condition $\Gamma = \Omega /\pi$. Above \qco{} the rapid increase of $\Gamma(Q)$ seems to saturate. However, this is the $Q$-region where the DHO model starts to fail in describing satisfactorily the experimental data, as previously shown in  Fig. \ref{IXSspectra}(f). The $Q$-dependence of the total inelastic intensity $I_{\rm inel}$, corrected for the atomic form factor of oxygen, is shown in Fig. \ref{DhoParameters}(c). At low $Q$, the Brillouin signal contributes a constant. The boson peak integrated intensity  should be proportional to $Q^2$ in the incoherent approximation \cite{Buchenau1985}. Extrapolating its measured integrated intensity to the low $Q$-value of Fig. \ref{DhoParameters}(c) leads to the thin curve drawn there. Below $Q$ = 2.5 nm$^{-1}$, the boson-peak contribution is thus a very small fraction of the spectral intensity. This confirms that what is seen in Figs. \ref{IXSspectra} and \ref{DhoParameters} is mostly produced by acousticlike excitations.

\section{Discussion}

In spite of the above evidence, claim was made during the IDMRCS conference \cite{ThisProc2005} that the rapid increase of the Brillouin linewidth {\em observed in the low-$Q$ region, below 2 nm$^{-1}$}, was just an artefact of the fitting procedure which did not take into account the growth in intensity of the transverse acoustic (TA) modes \cite{Ruocco2005}. This view is based on the idea that the polarization of acoustic modes could become ill-defined in a disordered medium at very short wavelengths. The signature of the transverse dynamics in the longitudinal spectra measured by means of IXS has never been established at the low-$Q$ values of interest here. Thus, we undertook such an analysis with the present high-accuracy data. The IXS spectra, up to $Q$ = 2.7 nm$^{-1}$, have been adjusted to a model including a supplementary DHO function for the TA modes. The ratio between the energy of the TA and LA modes was first kept constant, $\Omega_{\rm LA}/\Omega_{\rm TA}$ = 1.743, as found at lower frequencies with BLS \cite{Devaud1983}. With no surprise, the outcome of the analysis is a completely negligible TA-contribution over the entire $Q$-range analysed here. It amounts to less than 0.3\%, which probably means 0, with no noticeable effect on the linewidths found for the LA excitations. In a second round, the TA energies were let free, giving completely erratic results without any significance. A contribution of TA excitations can thus be discarded at $Q < q_{\rm co}$. This confirms that the inelastic signal seen in Fig. \ref{IXSspectra} is {\em definitely} due to the longitudinal acoustic modes {\em only} and, further, that the rapid increase of its linewidth {\em is neither an artefact nor a speculation}.

\begin{figure}
 \includegraphics[width=8.5cm]{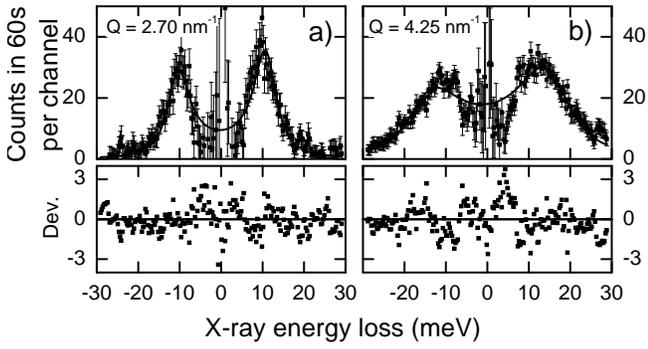}
\caption{IXS inelastic spectra of \dibo{} at 573 K above the crossover \qco{}. The {\em adjusted elastic part} is subtracted from both the experimental data and the global fit of each spectrum. The lines show the DHO adjustment in this global fit. The bottom panel of each subplot reports the difference between the experimental data and the fit normalized to the standard deviation at each point.}
\label{IXShighQspectra}
\end{figure}

We now turn to momentum transfers above \qco{}. We already mentioned that DHO fits start then to deviate significantly from the experimental data. This is illustrated in Fig. \ref{IXShighQspectra} with IXS spectra taken in \dibo{} at 2.7 and 4.25 nm$^{-1}$. The latter is actually the sum of four spectra accumulated on the next analyzer while taking measurements from 1.1 to 1.4 nm$^{-1}$. The additional small $Q$-dependence was taken into account in the fit which nevertheless is quite poor. A characteristic undulation develops in the deviation plots. It clearly shows that the DHO is no more a valid approximation. On the spectra, the dip near the origin, which is a signature of inhomogeneous broadening \cite{Vacher1999}, cannot be described at all by the DHO. Above \qco{}, the spectral shape should take into account the resonant coupling of rather localized boson-peak modes with the acoustic excitations. This leads to a rapid increase of the sound attenuation in this frequency range, in addition to an increased intensity derived from the extra modes. One observes in fact a progressive merging of the acoustic excitations with the modes forming the boson peak. The latter has been measured at the higher $Q$-values 23.4, 26.4, and 29.4 nm$^{-1}$. Its shape agrees then well with that found in Raman scattering \cite{Lorosch1984}, emphasizing the rather local nature of boson-peak modes. 

All these observations are in excellent agreement with previous ones on $d$-SiO$_2$ \cite{Ruffle2003}. Further, we have recently shown that the correspondence \OMco{} $\simeq$ \OMbp{} holds in most glasses investigated so far by IXS \cite{Ruffle2005}. This strongly supports the idea that hybridization of acoustic modes with the low-lying optic modes forming the boson beak is responsible for the observed inhomogeneous broadening of the acoustic modes. This hybridization is central to the boson-peak model of \cite{Gurevich2003}. We expect that in all glasses which present a sufficiently strong boson peak, there is a $Q$-region where $\Gamma$ increases rapidly. This also indicates that there is no physical reason to expect a universal law \cite{Matic2001b,Ruocco1999} $\Gamma \propto Q^2$, at least not below the crossover.

\begin{figure}
 \includegraphics[width=8.5cm]{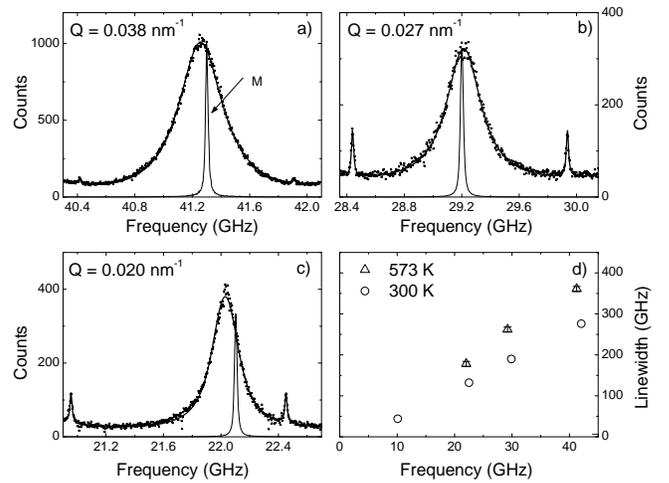}
\caption{(a)-(c) Brillouin light-scattering spectra of the LA-mode of \dibo{} and their fits to a DHO at 573 K for three scattering geometries. The small narrow peaks are produced by the elastic line. Their separation is the free spectral range.
M is a reference line showing the instrumental profile. It is obtained by electro-optic modulation of the laser near the Brillouin frequency. (d) Frequency dependence of the linewidth $\Gamma (\Omega)$ at two temperatures.}
\label{BLSspectra}
\end{figure}

The latter point is particularly well illustrated by the low-frequency sound attenuation of \dibo{} measured by means of optical Brillouin light scattering. Figs. \ref{BLSspectra}(a-c) present typical BLS spectra obtained at $T$ = 573 K in \dibo{} for three scattering geometries. Each plot shows the Stokes Brillouin signal at $\nu_{\rm S}$ between two orders of the elastic line, together with the modulation signal at $\nu_{\rm M}$. The Brillouin frequency and linewidth were measured also in function of the temperature. A very high accuracy is obtained on the Brillouin frequency thanks to the modulation line, the frequency of which is determined with the extremely high accuracy of a frequency generator. The instrumental resolution of the SFP used is FWHM $\sim$ 25 MHz. The spectrometer is then able to provide a 3 MHz (6 MHz) absolute accuracy on the frequency (linewidth) provided the Brillouin signal is sufficiently intense.
 
The frequencies and linewidths are extracted by adjustment to a DHO lineshape convoluted with the instrumental response, taking into account the broadening due to the finite collection angle. The frequency dependence of the Brillouin linewidth is illustrated in Fig. \ref{BLSspectra}(d) for the two temperatures studied in the IXS experiment. It clearly reveals a linear dependence of the sound attenuation with frequency. This behavior, $\Gamma \propto \Omega$, is characteristic of thermally activated relaxation processes at frequencies near the relaxation maximum, {\em i.e.} for $\Omega \tau \sim 1$, where $\tau$ is a typical relaxation time \cite{Hunklinger1976}. The same mechanism is also invoked to explain the ultrasound absorption in this glass \cite{Ciplys1981,Ruffle2005}. Thus in \dibo{}, the observed sound attenuation in the MHz and GHz ranges seems mainly controlled by thermally activated relaxation processes. Using the temperature dependence of the sound velocity, the anharmonic damping contribution \cite{Rat2005}, $\Gamma_{\rm anh} \propto \Omega^2$, is estimated to be about an order of magnitude smaller than the observed $\Gamma$. This is in contrast with $d$-SiO$_2$, or SiO$_2$ above 200 K, where the sound attenuation is mainly due to the anharmonic interaction of sound with the dominant thermal excitations forming the thermal bath \cite{Vacher2005}. Thus, the leading broadening mechanism strongly depends on the particular glass and on the measuring temperature $T$. It is only around the Ioffe-Regel crossover that a nearly universal behavior might be expected. 

\section{Conclusion}

To summarize, the existence of a rapid increase in the Brillouin linewidth measured in the THz region by means of IXS is definitely established in strong glasses. We found approximately $\Gamma \propto Q^4$, or $\Gamma \propto \Omega^4$, for the LA excitation. This increase seems to saturate around the boson-peak frequency where the Ioffe-Regel crossover is reached for the acoustic-like excitation, \OMco{} $\sim$ \OMbp. The most likely origin of this inhomogeneous broadening is the hybridisation of the sound wave with the low-lying optic-like modes forming the boson peak. Above the crossover, the inelastic spectra are strongly affected by this resonance and they cannot be described anymore by a simple DHO lineshape. The physical meaning of the DHO parameters extracted from the spectra above the crossover is questionable. At much lower frequencies, there exists in glasses several sound attenuation mechanisms whose strength depends both on the temperature and on the investigated glass. This can lead to a non-trivial frequency dependence which cannot be restricted to a single law $\Gamma \propto \Omega^2$. 

The authors thank A. Matic for the \dibo{} sample, R. Vialla for recent improvements to the high resolution BLS spectrometer, and J.M. Fromental for technical assistance during the IXS and BLS experiments.


\begin{thebibliography}{99}

\bibitem{Hunklinger1976} S. Hunklinger and W. Arnold, in {\em Physical Acoustics},
Vol. XII, W.P. Mason and R.N. Thurston Eds. (Academic Press, N.Y. 1976), p. 155.

\bibitem{Vacher1981} R. Vacher, J. Pelous, F. Plicque, and A. Zarembowitch, J. Non-Cryst. Solids {\bf 45}, 397 (1981).

\bibitem{Graebner1986} J.E. Graebner, G. Golding, and L.C. Allen, Phys. Rev. B {\bf 34}, 5696 (1986).

\bibitem{Buchenau1992} U. Buchenau, Y.M. Galperin, V.L. Gurevich, D.A. Parshin, M.A. Ramos, and H.R. Schober, Phys. Rev. B {\bf 46}, 2798 (1992).

\bibitem{Gurevich2003} V.L. Gurevich, D.A. Parshin, and H.R. Schober, Phys. Rev. B {\bf 67}, 094203 (2003).

\bibitem{Ioffe1960} A.F. Ioffe and A.R. Regel, Prog. Semicond. {\bf 4}, 237 (1960).

\bibitem{Rat1999} E. Rat, M. Foret, E. Courtens, R. Vacher, and M. Arai, Phys. Rev. Lett. {\bf 83}, 1355 (1999).

\bibitem{Ruffle2003} B. Ruffl\'e, M. Foret, E. Courtens, R. Vacher, and G. Monaco, Phys. Rev. Lett. {\bf 90}, 095502 (2003).

\bibitem{Courtens2003} E. Courtens, M. Foret, B. Hehlen, B. Ruffl\'e, and R. Vacher, J. Phys. Condens. Matter {\bf 15}, S1279 (2003).

\bibitem{Ruffle2005} B. Ruffl\'e, G. Guimbreti\`ere, E. Courtens, R. Vacher, and G. Monaco, arXiv:cond-mat/0506287 (2005).

\bibitem{Parshin2001} D.A. Parshin and C. Laermans, Phys. Rev. B {\bf 63}, 132203 (2001).

\bibitem{Foret2002} M. Foret, R. Vacher, E. Courtens, and G. Monaco, Phys. Rev. B {\bf 66}, 024204 (2002).

\bibitem{Zeller1971} R.C. Zeller and R.O. Pohl, Phys. Rev. B {\bf 4}, 2029 (1971).

\bibitem{ThisProc2005} Fifth International Discussion Meeting on Relaxation in Complex Systems, Lille, France, 2005.

\bibitem{Vacher1980} R. Vacher, H. Sussner, and M. v. Schickfus, Rev. Sci. Instrum. {\bf 51}, 288 (1980).

\bibitem{Matic2001} A. Matic, D. Engberg, C. Masciovecchio, and L. B\"orjesson, Phys. Rev. Lett. {\bf 86}, 3803 (2001).

\bibitem{Buchenau1985} U. Buchenau, Z. Phys. B {\bf 58}, 181 (1985).

\bibitem{Ruocco2005} G. Ruocco, communication at this Conference.

\bibitem{Devaud1983} M. Devaud, J.Y. Prieur, and W.D. Wallace, Solid States Ionics {\bf 9-10}, 593 (1983). 

\bibitem{Vacher1999} R. Vacher, E. Courtens, and M. Foret, Phil. Mag. B {\bf 79}, 1763 (1999).

\bibitem{Lorosch1984} J. Lor\"osch, M. Guzi, J. Pelous, R. Vacher, and A. Levasseur, J. Non-Cryst. Solids {\bf 69}, 1 (1984).

\bibitem{Matic2001b} A. Matic, L. B\"orjesson, G. Ruocco, C. Masciovecchio, A. Mermet, F. Sette, and R. Verbeni,
 Europhys. Lett. {\bf 54}, 77 (2001).

\bibitem{Ruocco1999} G. Ruocco, F. Sette, R. Di Leonardo, D. Fioretto, M. Krisch, M. Lorenzen, C. Masciovecchio, G. Monaco, F. Pignon, and T. Scopigno, Phys. Rev. Lett. {\bf 83}, 5583 (1999).

\bibitem{Ciplys1981} D. $\check {\rm C}$iplys, and J.Y. Prieur, J. Physique (Paris) {\bf 42}, C6-184 (1981).

\bibitem{Rat2005} E. Rat, M. Foret, G. Massiera, R. Vialla, M. Arai, R. Vacher, and E. Courtens, arXiv:cond-mat/0505558, 23 May 2005.

\bibitem{Vacher2005} R. Vacher, E. Courtens, and M. Foret, arXiv:cond-mat/0505560, 23 May 2005.

\end{thebibliography}
\end{document}